# Astrosociology: Interwiews about an infinite universe

Erik Høg
*Niels Bohr Institute, Juliane Maries Vej 30, 2100 Copenhagen Ø, Denmark*

If the universe is infinite now it has always been infinite. This is the opinion of many astronomers today as can be concluded from the following series of interviews, but the opinions differ much more than I had expected. Many astronomers do not have a clear opinion on this matter. Others have a clear opinion, but very different from the majority. Detailed arguments by two experts on general relativity are also included. Observations show that the universe is flat, i.e. the curvature is zero within the small uncertainty of measurements. This implies an infinite universe, though most probably we will never know that for certain. For comparison with the recent interviews, opinions during the past 2300 years since Aristotle about the universe being finite or infinite have been collected from literature, and it appears that the scientists often had quite definite opinions. © Anita Publications. All rights reserved.

## 1 Introduction

"If the universe is infinite now, then it has always been infinite", is this correct? or what do you think? That was my question to many astronomers during a conference in 2012 and the many different answers are subject of the present report.

You are absolutely right, Erik, was the answer from Pavel Naselsky when I asked him several years ago. Pavel has been working on cosmology for many years with Igor Novikov in Copenhagen, so I knew he would be the right person to ask. But I have also received quite different answers, one of them from a very respected astronomer who knew the cosmologists's world from inside. He said the dominance today of the big bang model of the universe had come along with suppression of other views, sometimes even with mafia-like methods.

I am myself not a professional cosmologist, but I try to tell about cosmology to ordinary people and my answers must be correct to the best of astronomy today. I personally believe the big bang model is essentially correct and I explain the expanding universe as follows.

## 2 The expanding universe

The universe contains the Earth, the Moon and planets, stars and galaxies, and gas and dust between them, and all the radiation. In short, all we can think of, matter, space and radiation belong to the universe, constitute the universe. According to modern cosmology 96 percent of the matter in the universe is invisible does not consist of atoms but dominates gravitation in the universe. This matter is called dark matter and dark energy, but we do not yet know what it really is. It may be exotic particles, but it is certainly not atoms.

We suppose that the universe is roughly the same everywhere and looks similar in all directions, seen from any place in the universe. This is the "cosmological principle", astronomers saying that the universe is homogenous and isotropic. There are good reasons for this assumption from the part of the universe which can be observed, but the experts have also good theoretical reasons to say that the universe much further away may be very different. This is due to the "inflation", an enormous expansion that took place when the universe was about $10^{-35}$ seconds old.

The expression "expansion of the universe" is used by astronomers for the phenomenon that the distance (here taken as the "coordinate distance") to galaxies in distant galaxy clusters increases with time in proportion to the distance to the cluster, and this can be directly observed. After a very long time all distances have become, e.g., three times larger. But this does NOT mean that the galaxies rush through space, i.e. through the universe. The phenomenon is understood as due to an expansion of the space itself,

---

*Corresponding author :*
*e-mail: Erik.hoeg@get2net.dk* (Prof Erik Høg)

of the universe itself, while the galaxies hardly move at all in space. This is what the theory of general relativity says, and we observe in fact that even the wavelength of light increases with time during the passage through space. – A common misrepresentation is that the expansion was due to an explosion which it is not. Nor should one say that the universe began as very small, truly it was very dense and hot, but not necessarily small, it might even have been infinite.

One may imagine a part of the universe being bread dough with raisins where the dough is the space and the raisins are galaxy clusters. As the dough rises the distances between raisins increase, corresponding to the distances between clusters of galaxies becoming larger when the universe expands. But a raisin does not increase because it is held together by the much stronger forces among molecules, and in the universe a cluster of galaxies does not expand because it is held together by the force of gravitation from the mass of the galaxies. Similarly, a galaxy or our solar system does not expand with the universe.

**3 Infinity and distances in cosmos**

It would be interesting to compare the answers to my question about an infinite universe with the opinions collected from groups of astronomers at other times in history, but I am not aware of any such collection. For comparison I can therefore only use authoritative answers found in astronomical textbooks. I have consulted four textbooks Dreyer (1905), Bondi (1952), van Helden (1985), and Unsöld (1967) as well as papers by Bonnor (1964) and Richter (2001) from which I briefly extract.

Aristotle (384-322) argues very definitely (Dreyer p.109) that "…the material universe cannot be infinitely extended, since a line from the earth's centre to an infinitely distant body could not perform a complete circular revolution in a definite time (twenty-four hours); and as there cannot be bodies at an infinite distance, neither can space be infinite, since it is only a receptable of bodies. The heavens are without beginning and imperishable, since they cannot be the one without the other, while Plato (427-347) had supposed that though the world had been created, it would last forever."

Herakleides of Pontus or Heraklides Ponticus (~390 to after 322) "clearly and distinctly taught that it is the earth which turns on its axis from west to east in twenty-four hours", according to Dreyer p.123. "…He called the world a god and a divine mind. … We are also told that he considered the Cosmos to be infinite, and that he considered each planet to be a world with an earth-like body and an atmosphere."

Most probably, the view of Aristotle was shared by all astronomers up to Kopernicus (1473-1543) who asserted that the earth rotates in 24 hours, and that the universe does not rotate at all. But it is known that Tycho Brahe (1546-1601) shared the belief in the 24 hour rotation period of the universe, the earth staying without rotation.

Ptolemy in the second century of our era, gave distances to the Moon, Sun, planets, and the stars, partly obtained from his predecessors and as given by van Helden (p.27). All these values were accepted by astronomers, including the Islamic, up to Johannes Kepler (1571-1630) who was the first to express doubt about the distance to the Sun which was in fact 20 times too small as it turned out from research during the subsequent century. The distance to the stars, all situated on a sphere centred on the earth, was 20000 earth radii according to Ptolemy. This radius translates into 0.000 014 light years, see Table 1.

The Table 1 shows the radius of the visible universe, defined as the largest distance from which light could be received on earth according to astronomers' estimate. The last column gives the authority. The last line gives what is often called the cosmic horizon, obtained as the age of the universe multiplied with the velocity of light, but this requires some explanation. The distance to the last scattering surface which released the radiation now observed as the microwave background, CMB, is very different. This distance was 42 million light years when the radiation was released 380 000 years after big bang, and our CMB photons, i.e. those destined for us, in fact moved away from us with a speed about 50 times the velocity of light. But with time they began to approach us. The atoms at this surface have since moved with the expansion of the universe to a distance 1100 times larger. The atoms have taken part in the evolution to become stars and galaxies, now at a distance of 45 billion light years. Outside this horizon the universe is much larger, perhaps infinite, but it is similar on large scales to the visible universe, according to the widely accepted cosmological principle. – Otherwise, the table is selfexplanatory.

Tabel 1. Radius of the visible universe, according to authorities during the past 2000 years.

| Year | Radius – light years | Objects | Sources |
|---|---|---|---|
| 150-1600 | 0.000 014 | Sphere of stars | C. Ptolemy |
| 1800 | 3 000 | Milky Way | W. Herschel |
| 1900 | 30 000 | Milky Way | J.C. Kapteyn |
| 1920 | 200 000 | Globular clusters | H. Shapley |
| 1930 | 30 000 000 | Galaxies in Virgo | E. Hubble |
| 1960 | 2 000 000 000 | Most distant galaxies | Palomar 5 meter |
| 2003 | 13 700 000 000 | Visible universe | Hipparcos, Hubble, WMAP |

Bondi (1952) discusses the observational basis for cosmology and the many cosmological theories of his time. The names are: the Bondi-Gold and Hoyle formulations of the steady-state theory, Milne's kinematic relativity, Einstein's model, Eddington's theory, the Lemaitre-Eddington model, Jordan's theory, and Dirac's theory, a bewildering multitude to which most astronomers paid only little attention, including myself as a young student. George Gamow's big bang theory is not mentioned, it had just about been proposed when Bondi finished his book in October 1950, and the name "big bang" had only been coined the year before, by Fred Hoyle on 28 March 1949.

Infinity of both time and space occurs in several models, and infinity is a subject in Bondi's chapter III: "The background light of the sky". The chapter is centred on a discussion of Olbers' paradox: "…pointed out as early as 1826 in a remarkable paper by Olbers", a paradox of "deep significance for cosmology."

The Olbers' paradox is subject of a paper by Bonnor (1964). The author writes: "The Olbers' paradox is roughly the following. If we lived in an infinite static universe homogenously populated with stars, the night sky would be brighter than the present day sky. It would in fact be everywhere as bright as at the surface of a star. Although this was a paradox when Olbers discovered it in 1826, it is not so today because we know that the universe is not static and does not consist of a homogenous distribution of stars." Bonnor derives a formula for the flux of light at a point in a general cosmological model and shows that of all plausible models, none shows a conflict with the observed light-flux, not even in oscillating models. Thus, the paradox cannot be used to distinguish between the models.

Richter (2001) discusses Olbers' paradox in an historical context. The paradox became a subject for cosmology only about 1950 when Bondi drew attention to it as mentioned above. The paper by Wilhelm Olbers (1781-1862) from 1823 appeared 1826 in three languages, German, English, and French but they were hardly noticed at that time. For Olbers the issue was the transparency of space as the title says, not at all cosmology since he accepted without any discussion the apodictic claim by Immanuel Kant (1724-1804) about the infinity of space and time. Olbers' solution to the problem was that light was absorbed by a very thin medium in space, sufficient to resolve the paradox.

This solution could not hold later on when the theory of radiation and energy became known because the medium would be heated and glow. The energy cannot disappear. A solution was given by Bonnor, see above, and we could add the argument that the universe has a history, a finite age of about 13 billion years would be sufficient to make the paradox disappear.

Unsöld (1967) presents a description of astronomy for a general audience. On p.3 he mentions infinity: "Only in our time do we begin to appreciate the deep meditations of cardinal Nicolaus Cusanus (1401-1464). It is very interesting to see how ideas about the infinity and the world and ideas about quantitative studies of nature for him originate from the religious and theological thinking." - (My translation from German.)

Unsöld discusses the Olbers paradox on pp.314: "H.W.M. Olbers was one of the first astronomers to consider a cosmological problem from the empirical point of view."

Kepler is probably the first to wonder about the darkness of the night sky, in 1606, and he takes this as a clear evidence for a finite cosmos (Richter p.105). He uses this as an argument against speculations by William Gilbert and Giordano Bruno.

Carl Fredrich von Weizsäcker (1912-2007) shall finally be quoted from his lectures "Die Geschichte der Natur" held in Göttingen in 1946, published in 1962. According to Richter p.112 and here briefly and freely translated: "Few questions in science are further away from the immediate needs of anyone than such a question as the infinity of the universe. But few questions can give rise to more heated discussion. Central for an understanding of the answers is to realize that human attitudes, human types appear, all have their own ways to answer. The person tries to understand nature objectively, but coming to the deepest issues, he unexpectedly sees himself as in a mirror." How true that may be should appear from the following interviews.

Von Weizsäcker is the most perfect speaker I ever heard. I attended his weekly lectures on philosophy in Hamburg in 1959 in an overfilled Auditorium Maximum, students sitting on the floor and stairs. He spoke for 45+45 minutes without manuscript or pictures, beginning perhaps writing four or five words on a small blackboard, and his speech was ripe for printing.

**4 An infinite universe**

Returning now to August 2012, I asked the question about an infinite universe many times during the conference of 2000 astronomers in Beijing. The question often served as a good starting point for a discussion in the corridor or during a bus tour, and the answers are together a kind of sociology, 'astrosociology' it may be called; or is it 'astropsychology'?. I should add that the answers are recorded as given without telling the answers already given by others. Some answered immediately, a few were reluctant, especially to let their answer become public, because of the risk for misunderstandings. But I could often convince them of the importance to do so when I tried.

One of the first days of the conference, I heard the talk by Wendy Freedman, an overview of the cosmic distance scale, given for an audience of over 400 astronomers, all early risers since it began at 8:30 in the morning. When it came to questions an hour later, I asked for a microphone which I do not often do in the presence of so many people. I asked: "At the beginning of your talk you showed a slide with the universe before big bang (the slide contained a small black spot on a white screen) and it looked quite small. But now comes my question: If the universe is infinite now then it always was infinite, at inflation, even before inflation, even back at the Planck time. Is this correct?"

Her answer came with a smile: "Yes, you are right", and she continued to explain that the observable universe is much smaller when we go back in time because all distances are smaller, and because the time light or signals can have travelled is shorter as we approach the big bang.

Soon after, I asked Gustav Tammann, the expert on the expansion of the universe, Albert Einstein medal in 2000, and his answer came immediately: "Yes, of course. It is a purely mathematical question." There was no time for discussion, but the next day I met him again and said: "No, it is not purely mathematical, because you can easily have a function of time in mathematics with a discontinuity. It goes from a finite value to infinity, but you cannot have that in a physical system. The universe cannot suddenly go from a finite size to infinity. That would require infinite velocities which is impossible." He immediately accepted this and said: "So I was right in the matter but not in the reasoning!"

Another said: "Yes, it is simply logical." I replied that it is not just a logical or mathematical necessity, the physical nature must be taken into account. But he did not agree on that.

Sergei Klioner, expert on general relativity, said: "What do you mean by infinite?" I wondered about that question from Sergei whom I have known for many years. But I explained the best I could what I meant by infinite space in three dimensions, and he agreed that the universe has three spatial dimensions and it expands with time, which is a fourth dimension but not spatial. Coming back to my original question, he did not agree that the universe has necessarily been infinite at any earlier time even if it is infinite now. He answered in a manner that I later thought he had misunderstood me. So, I asked him again some days later and then understood that he had completely understood my question, but he had an opinion which was so interesting that I asked him to write to me, and he agreed to do so. His answer follows below.

Jean-Claude Pecker, now 90 years of age and still very active, immediately agreed that the universe has always been infinite and he even added that it has infinite age. He has written a paper with Jayant Narlikar and others that the cosmic microwave background, CMB, comes from dust in our galaxy, and not from an afterglow of any big bang. Pecker later added by mail: "What does that mean, an "infinite Universe in space and time"? Not a result of measures, of course: Noone has ever measured an "infinite" or a "null" quantity! The only thing that one may say is: Whatever the object $A$ that one can observe and perhaps study, I am convinced there is another object $B$ further away, at a larger distance than the object $A$, and that I cannot (so far) observe and study. Whatever the epoch $T$ at which my observations and studies can reveal some reality, I am convinced there is another epoch $S$, earlier than $V$, that I could observe and study, or perhaps that I shall never be able to observe and study." That is what I mean by "believing" in the infinity of space and time.

Edward van den Heuvel, high-energy astrophysicist, answered my email: "Dear Erik, This is a very good question. I think that indeed the answer is that if it is infinite now, it must always have been infinite. I can see no other way."

Kristian Pedersen, head of the Danish National Space Institute, wrote that he thinks the universe is in fact infinite now and that it has always been infinite.

Most others also agreed, but some did not want to draw any conclusion although they saw the problem of a discontinuity if the universe switched from finite to infinite, "but what do we know about time in the universe?" one asked. Many did not have an outspoken opinion. Some had obviously never thought of such a question and some seemed to worry of the small universe at the time of big bang and about the possibility of multiverses. All together, I spoke to about twenty astronomers or corresponded by email.

One of the answers is of interest because it can represent several similar ones, but it does not agree with the modern view of the universe explained in the introduction. It does not obey the cosmological principle since the answer says that there is infinite empty space outside the "Universe" where all matter is located. The mail said: "The Universe cannot be infinite because the matter and energy it contains must be finite. Infinity is only a mathematical concept, nothing more and nothing real can be infinite. Space is for sure infinite and the Universe as defined by the matter (visible or dark) and the energy it is composed of expands into this infinite space." It continued later: "I of course respect the opinion of professional cosmologists, however, I am convinced that when it comes to these matters their opinions are as useful or useless as, say, those of any other (mathematically trained) person that is no expert in the field: We are completely clueless about the nature of the Universe and its extent."

Another astronomer had opinions about the structure of the universe which were so unconventional that he did not want me to include them here because he suspected he would be recognized and have problems in his career.

Peter Naur replied immediately that my question could hardly be answered at all. But I would like to quote his later answer by letter since I rate Naur and his opinions highly; he was my astronomy mentor in the 1950s, he became the first Danish professor of computer sciences, he won the 2005 Turing-award, also known as the "Nobel-prize of computing science", and he is now working on psychology writing e.g. "An anatomy of human mental life" in 2005. Naur answered, translated from Danish: "Your question about the universe and infinities I consider as harmful philosophical nonsense. Of the same character as when the famous Hawking says that the human brain is a computer. Such nonsense has now since 1958 under the name of cognitivisme been a poison for psychology. It is harmful because cognitivism will prevent us from ever understanding how mental life goes on in the nervous system."

In the course of the interviews I have noticed that some people cannot even think quietly of an infinite universe. My advice is just to use the infinite series of natural numbers 1, 2, 3… and then to think of so many million light years. In this way it is simple to think of infinity and to understand it without confusion. But confusion comes when people immediately start to discuss other questions: How could the universe begin as a singularity if it is infinite now?? How can you prove that it is infinite?? If you travel even with the speed of light and the universe expands…?? etc. etc. I say stop! Take one question at a time.

Some people have strong feelings against an infinite universe, and even against thinking of it as such. This amazes me since it should make no difference to us human beings whether the universe is infinite or just immensely large as it surely is.

Here follows the correspondence with two experts on general relativity (GR). I found it important to include these in order to give an idea of the different views and ways of expression among experts.

**Sergei Kopeikin -** A correspondence on 1 Sep. 2012:

SK: Regarding your question I should say it is not simple to answer. In addition, the answer is not unique. It depends on a particular model of the universe and on a particular set of observers. I am also guessing that you are probably asking about the spatial cross-section of the universe that is a (hyper)surface of constant time. There is also a concept of the cosmological horizon of particles. It can be finite at the given epoch while the space of the universe itself is infinite. There are three types of the Friedmann universe – closed, flat and open. The flat and open universes are infinite but they originated from a singularity that is at the instant of the Big Bang the universe was just a single point! It sounds like physical absurd but mathematically it is possible. Many physicists do not like the initial singularity and they invented various models which preclude the original singularity but the universe in the original state is finite in any case. Well, to make a long story short my answer is that I do not share the opinion that if the universe is infinite now it was always infinite. There are too many mathematical tricks which can be played to comfort my opinion. On the other hand, the opposite point of view is also possible. Dialectics!

EH: My reasoning was very simple minded. But precisely where am I wrong??? Can there be infinite velocities in a physical universe???

SK: I agree that infinitely fast jumps from place to place requires an infinite velocity, and this is unphysical. I would not accept a system that had such a property. On the other hand, the infinite velocity can be reached by a particle after it falls down behind a horizon of a black hole. Such a particle inevitably will approach the central singularity (which is a physical singularity of a Riemann tensor = curvature) and as it approaches the singularity, its speed approaches to infinity. It takes a finite time, by the way, as measured by the clock on the particle. This is what mathematical solution says. Nonetheless, we have to admit the existence of the central singularity, and nobody knows what exactly is going on in there. So, you may consider this example as unphysical though Hawking and Penrose would definitely argue against such an opinion. They worked out a number of theorems on singularities and proved their existence in general relativity. The subject is pretty complicated though.

EH: You speak of the singularity of a black hole. You mentioned this in our talk of the universe perhaps because you think there was or could have been similar conditions close to the big bang? But if the universe is infinite now and you do not go back too close to big bang, the universe was probably infinite then? Could it be infinite just after inflation? or before inflation? or even at the end of Planck time?

SK: I believe it could. It depends on the structure of the original spacetime manifold. There are "crazy" ideas about extra dimensions in the context of the string theory, quantum gravity theory, etc. The universe at that time could be totally different from the standard 4-dimensional spacetime manifold where we "see" only 4 dimensions with the other dimensions being rolled up, and invisible today. When we say "infinite" probably it means the time taken by light to go across the universe. In those manifolds at the early stages of the universe the light ray has no "time" to cover the entire manifold – so we should interpret it as infinite. The question is rather tricky, and requires precise mathematical definitions and operations. As I mentioned previously, the answer may depend on our postulates. Unfortunately, we have no experimental data about the early stage of the universe. Hence, we can only guess and believe in one or another scenario.

**Sergei Klioner -** From a correspondence in September 2012, agreed with Klioner to be included here:

First of all I should tell you that I do not think that it is a good idea to put our discussion in a written form somewhere. It depends, of course, but such semi-philosophical discussions "signed" by "experts" could be easily misinterpreted by general public and lead to some wrong impressions. Moreover, I do not feel myself as "expert" in this discussion. As I said to you, my comments are not "scientific", but more personal if you wish. Anyway, I trust you. So my position is:

1. Our current knowledge of cosmology suggests that the Universe is infinite. However, this statement can evolve as soon as new observational data arrive or as soon as we invent new ways to interpret our current observational data. For me it means that the fact that the Universe is infinite cannot be considered as well established. It is simply our current understanding of the data and theory.

2. If we assume that the Universe is infinite now (e.g. described by the flat Robertson-Walker metric in the framework of general relativity), then it was infinite in all those "moments of time" at which general relativity (or an alternative metric theory of gravity) correctly describes the nature. I do not know when General Relativity (or an alternative metric theory of gravity) started to be adequate. I also do not have enough knowledge to speculate about this.

3. Personally and absolutely not scientifically I do believe in the existence of absolutely different physics. For example, the world would be so much boring for me if it would be not possible to create a technical device transporting me from the Earth to a planet in the Andromeda galaxy within a second. From this point of view, I am not inclined to extrapolate our current physical knowledge (based on the kind of observations we can do now) to some other regimes and attempting to answer some global semi-philosophical questions like if the Universe is infinite in reality or not.

Once again: actually I would not be really happy if all this appears in written form in a place accessible for the general public. – But soon after, Klioner has agreed to let it become public, although reluctantly.

## 5 My personal opinion

I have been urged, by Pavel Naselsky, to include my own opinion, not only to ask others. My personal opinion is rather conventional, I trust the cosmologists when they say that the big bang theory is a good basis for a consistent description of the universe and its evolution through 14 billion years. The evolution can be followed from an age of a very small fraction of a second, perhaps $10^{-35}$ s. Before that time better than just "intelligent guesses" are available, see Hartle & Hawking (1983).

The inflation theory, advocated by, e.g., Alan Guth and Andrei Linde since 1980 says that the entire universe is very much larger than our observable part within the cosmic horizon. To say "very much larger" is quite weak compared to what Linde (2007: p131) says: "Even if the initial size of the inflationary Universe was as small as the Planck scale, $l_\mathrm{P} \sim 10^{-33}$ cm, one can show that after $10^{-30}$ s of inflation this acquires a huge size of $l \sim 10$ to the power of 1000 billion cm. This number is model-dependent…" Other numbers are found in literature: at least a factor $10^{78}$ for the volume. The inflation explains that the observable universe is very nearly flat and as isotropic and homogenous as observations show.

The cosmic horizon is defined by the matter from which the cosmic background radiation was emitted when the universe was 380 000 years old and it has presently a distance of about 45 billion light years (proper distance). Out there, much beyond this horizon, the universe or 'multiverse' is probably very different from our 'small' part. It is even expected that the laws of physics are different from ours in some parts of the multiverse.

I would like to end these discussions about infinity of the universe by again quoting Carl Fredrich von Weizsäcker as at the end of Section 2: "Few questions in science are further away from the immediate needs of anyone than such a question as the infinity of the universe. The person tries to understand nature objectively, but coming to the deepest issues, he unexpectedly sees himself as in a mirror."

***Acknowledgements:*** I am grateful to all those answering my question whatever their answer was and for their permission to quote, and also to Virginia Trimple for comments to a previous version of this paper which led to an entertaining hunt for infinities on my own bookshelves.